\documentstyle[epsfig]{aipproc}
\def\etal{{\it et al.}}

\begin{document}
 
\title{Linear State Space Modeling of Gamma-Ray Burst Lightcurves}
\author{David Band$^{1}$, Michael K\"onig$^{2}$ and Anton Chernenko$^{3}$}
\address{$^{1}$CASS, UC San Diego, La Jolla, CA  92093\\
$^{2}$University of T\"ubingen, Germany\\
$^{3}$Space Research Institute, Moscow, Russia}

\maketitle
\begin{abstract}
Linear State Space Modeling determines the hidden autoregressive (AR)
process in a noisy time series; for an AR process the time series'
current value is the sum of current stochastic ``noise'' and a linear
combination of previous values.  We present preliminary results from
modeling a sample of 4 channel BATSE LAD lightcurves.  We determine
the order of the AR process necessary to model the bursts. The
comparison of decay constants for different energy bands shows that
structure decays more rapidly at high energy. The resulting models can
be interpreted physically; for example, they may reveal the response
of the burst emission region to the injection of energy. 
\end{abstract}
\section*{Introduction}
Hidden in BATSE's superb gamma-ray burst lightcurves in different
energy bands are temporal and spectral signatures of the fundamental
physical processes which produced the observed emission.  Various
techniques have been applied to the BATSE data to extract these
signatures, such as: auto- and crosscorrelations of lightcurves in
different energies\cite{band97}; Fourier transforms\cite{shaviv96};
lightcurve averaging\cite{mitrofanov96}; Cross-Fourier 
transforms\cite{kouveliotou92} and pulse
fitting\cite{norris96}.  Here we propose to use Linear State Space
Models (LSSM) to study the gamma-ray burst lightcurves. 
%

LSSM estimates a time series' underlying autoregressive (AR) process
in the presence of observational noise.  An AR process assumes that
the real time series is a linear function of its past values
(``autoregression'') in addition to ``noise,'' a stochastic component
of the process.  Since the noise adds information to the system, it is
sometimes called the ``innovation''\cite{scargle81}. A moving average
of the previous noise terms is equivalent to autoregression, and
therefore these models are often called ARMA (AutoRegressive, Moving
Average) processes\cite{scargle81}.  While ARMA processes are
simply mathematical models of a time series, the resulting model can
be interpreted physically, which is the purpose of their application
to astrophysical systems.  For example, the noise may be the injection
of energy into an emission region, while the autoregression may be the
response of the emission region to this energy injection, such as
exponential cooling. 

The application of LSSM to burst lightcurves can be viewed as an
exploration of burst phenomenology devoid of physical content: how
complicated an AR process is necessary to model burst lightcurves? Can
all bursts be modeled with the same AR process?  However, because
different types of AR processes can be interpreted as the response of a
system to a stochastic excitation, characterizing bursts in terms of
AR processes has physical implications.  Since we have lightcurves in
different energy bands, we can compare the response at different
energies.  For example, the single coefficient in the AR[1] process 
(the nomenclature is described below) is
a function of an exponential decay constant.  If the lightcurves in
all energy bands can be modeled by AR[1] then we have decay constants
for every energy band.  Since most bursts undergo hard-to-soft
spectral evolution\cite{ford95,band97} and temporal structure is
narrower at high energy than at low energy\cite{fenimore95}, we expect
the decay constants to be shorter for the high energy bands. 
\section*{Linear State Space Models}
The purpose of the LSSM methodology is to recover the hidden AR
process.  If the time series $x(t)$ is an AR[p] process then 
\begin{equation}
x(t) = \sum_{i=1}^p a_i x(t-i) + \epsilon(t,\sigma_x^2)
\end{equation}
where time is assumed to advance in integral units.  The ``noise'' (or
``innovation'') $\epsilon(t,\sigma_x^2)$ is uncorrelated and possesses
a well-defined variance $\sigma_x^2$; the noise is usually assumed to
be Gaussian.  Since the burst count rate cannot be negative, we expect
the noise also cannot be negative. A Kolmogorov-Smirnov test is used
to determine when p is large enough to model the system
adequately\cite{honerkamp93,koenig97}. 

If p=1, the system responds exponentially to the noise with a decay
constant $\tau$, and 
\begin{equation}
a_1 = e^{-1/\tau} \quad .
\end{equation}
The p=2 system is a damped oscillator with period $T$ and
relaxation time $\tau$, 
\begin{equation}
a_1 = 2\cos \left( {{2\pi}\over{T}}\right) e^{-1/\tau} \quad
   \hbox{ and }\quad a_2 = e^{-2/\tau} \quad . 
\end{equation}
Thus, the lowest order AR processes lend themselves to obvious
physical interpretations. 

Unfortunately, we do not detect $x(t)$ directly, but a quantity $y(t)$
which is a linear function of $x(t)$ and observational noise: 
\begin{equation}
y(t) = C x(t) + \eta(t,\sigma_y^2)
\end{equation}
where in our case $C$ is an irrelevant multiplicative factor and
$\eta$ is a zero-mean noise term with variance $\sigma_y^2$; $\eta$ is
also often assumed to be Gaussian.  The LSSM code uses
the Expectation-Maximization algorithm\cite{honerkamp93,koenig97}. 
\section*{Application to Bursts} 
We have thus far applied our LSSM code\cite{koenig97} to 17 gamma-ray 
bursts. We used the 4-channel BATSE LAD discriminator lightcurves 
extracted from the DISCSC, PREB, and DISCLA datatypes, which have 
64~ms resolution; the energy ranges are 25--50, 50--100, 100--300 
and 300--2000~keV.  Each channel was treated separately, resulting 
in 68 lightcurves.  Of these lightcurves, 52 could be modeled by AR[1], 
13 by AR[2] and 3 by AR[4].  Thus there is a preference for the 
simplest model, AR[1].  Note that Chernenko \etal\cite{chernenko98}
found an exponential response to a source function in their soft
component.

\begin{figure}[b!] 
\centerline{\epsfig{file=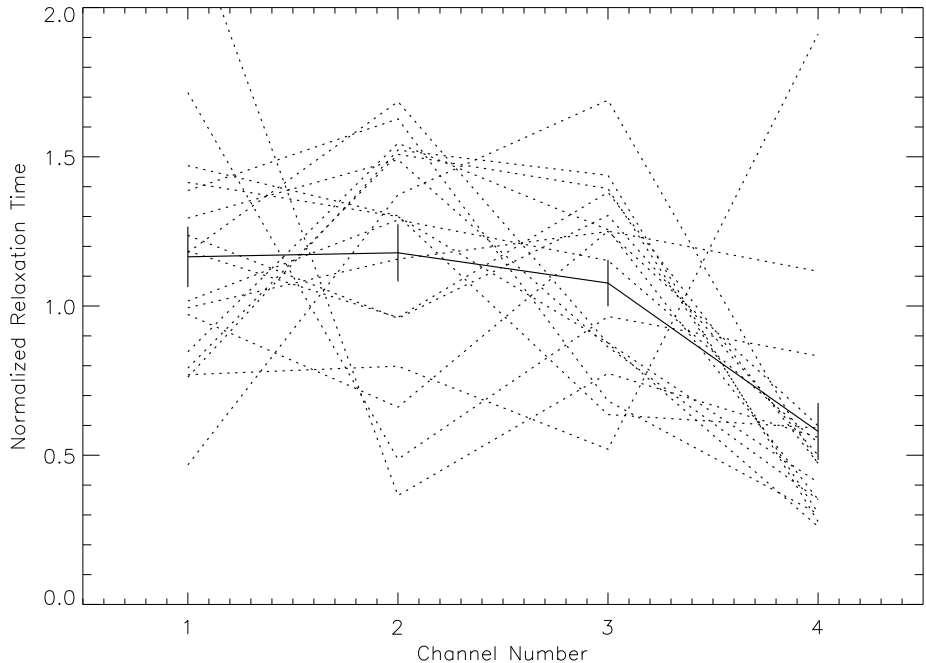,width=5.25in}}
\caption{AR[1] time constants by channel, normalized by the
average over all channels.  The dotted curves are for each of the 17
bursts, while the solid curve is the average. Note that the time
constants decrease from channel 1 to channel 4, as expected for
hard-to-soft evolution, although there is a great deal of scatter for
the individual bursts.}
\label{fig1}
\end{figure}
Figure~1 presents the normalized relaxation time constants for
the bursts in our sample, as well as their average.  Even for models 
more complicated that AR[1] a relaxation time constant can be identified.  
As expected, the averages of these time constants become shorter as 
the energy increases from channel~1 to channel~4, consistent 
with the trend found in quantitative studies of spectral 
evolution\cite{band97,ford95} and the qualitative inspection of 
burst lightcurves. 

\begin{figure}[b!] 
\centerline{\epsfig{file=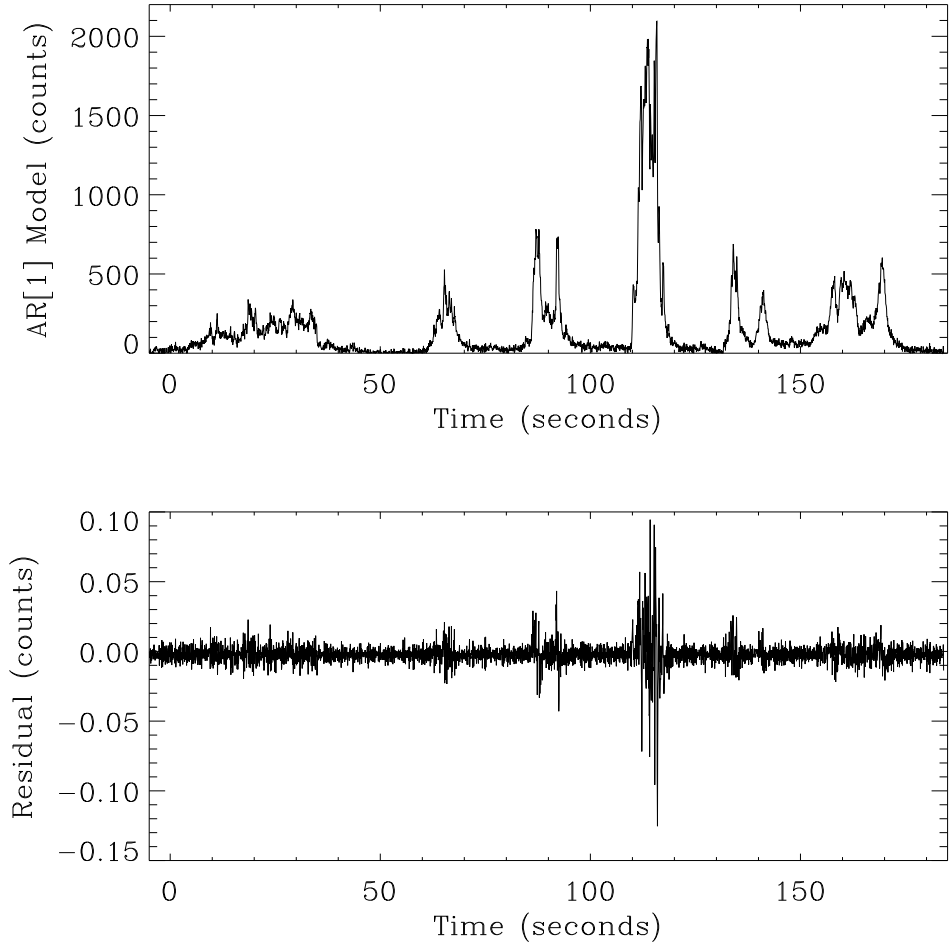}}
\caption{AR[1] model for channel~2 (50-100 keV) of GRB~940217 (top) 
and the residuals ($\eta$ in
eq.~[4]) between the data and the model (bottom).} 
\label{fig2}
\end{figure}
In Figure~2 we present the analysis of GRB~940217, the 
burst with an
18~GeV photon 90 minutes after the lower energy gamma-ray emission
ended\cite{hurley94}. 
As can be seen, the
residuals are much smaller than the model and are consistent with
fluctuations around 0; plots for the data and the model are
indistinguishable, and only one is presented.  The amplitude of the
residuals increases as the count rate increases (attributable in part
to counting statistics), but there is no net deviation from 0.

\section*{Future Directions}
We plan to  apply the LSSM code to a large number of bursts.  We will 
compare the order of the underlying AR process and the resulting 
coefficients obtained for the different energy lightcurves of the 
same burst and for different bursts.  In this way we can search for 
hidden classes of bursts and explore the universality of the physical 
processes. 

The ``noise'' $\epsilon(t)$ might be a measure of the energy
supplied to the emission region (although which physical processes
are the noise and which the response is model dependent).  
Therefore characterizing $\epsilon(t)$ may probe a deeper level of the
burst phenomenon.  The $\epsilon(t)$ lightcurves for the different
energy bands should be related; we expect major events to occur at the
same time in all the energy bands, although the relative intensities
may differ. 

Many of the bursts consist of well-separated spikes or complexes of
spikes.  We will apply the LSSM code to each part of the burst to 
determine whether the same order AR process characterizes the entire 
burst, and if so, whether the AR process has the same coefficients.  
This will test whether the physical processes remain the same during 
the burst.
\section*{Acknowledgments}
D.~Band's gamma-ray burst research is supported by the {\it CGRO} guest 
investigator program and NASA contract NAS8-36081. 

\end{document}